\documentclass[fleqn,10pt]{wlscirep}
\usepackage[utf8]{inputenc}
\usepackage[T1]{fontenc}

\usepackage{graphicx}
\usepackage{nameref}
\usepackage[per-mode=symbol, exponent-product=\times]{siunitx}
\title{Slow and fast light behavior of single photons from a quantum dot interacting with the excited state hyperfine structure of the Cesium D$_1$-line}

\author[1,*]{Tim Kroh}
\author[2]{Janik Wolters}
\author[1]{Andreas Ahlrichs}
\author[3]{Andreas W. Schell}
\author[4]{Alexander Thoma}
\author[4]{Stephan Reitzenstein}
\author[5]{Johannes S. Wildmann}
\author[6,7]{Eugenio Zallo}
\author[8]{Rinaldo Trotta}
\author[5]{Armando Rastelli}
\author[7]{Oliver G. Schmidt}
\author[1]{Oliver Benson}

\affil[1 ]{Department of Physics, Humboldt-Universit\"at zu Berlin, 12489 Berlin, Germany}
\affil[2 ]{Department of Physics, University of Basel, 4056 Basel, Switzerland}
\affil[3 ]{CEITEC Brno University of Technology, 621 00 Brno, Czech Republic}
\affil[4 ]{Institute of Solid State Physics, Technische Universit\"at Berlin, 10623 Berlin, Germany}
\affil[5 ]{Institute of Semiconductor and Solid State Physics, Johannes Kepler Universit\"at Linz, 4040 Linz, Austria}
\affil[6 ]{Paul-Drude-Institut für Festk\"orperelektronik, 10117 Berlin, Germany}
\affil[7 ]{Institute for Integrative Nanosciences, Leibniz IFW Dresden, 01069 Dresden, Germany}
\affil[8 ]{Department of Physics, Sapienza University of Rome, 00185 Rome, Italy}

\affil[* ]{ Corresponding author: tim.kroh@physik.hu-berlin.de}

\begin{abstract}
Hybrid interfaces between distinct quantum systems play a major role in the implementation of quantum networks. 
Quantum states have to be stored in memories to synchronize the photon arrival times for entanglement swapping by projective measurements in quantum repeaters or for entanglement purification. 
Here, we analyze the distortion of a single photon wave packet propagating through a dispersive and absorptive medium with high spectral resolution. 
Single photons are generated from a single In(Ga)As quantum dot with its excitonic transition precisely set relative to the Cesium D$_1$ transition. 
The delay of spectral components of the single photon wave packet with almost Fourier-limited width is investigated in detail with a 200~MHz narrow-band monolithic Fabry-P\'erot resonator. 
Reflecting the excited state hyperfine structure of Cesium, ``slow light'' and ``fast light'' behavior is observed. 
As a step towards room-temperature alkali vapor memories, quantum dot photons are delayed for 5~ns by strong dispersion between the two 1.17~GHz hyperfine-split excited state transitions. 
Based on optical pumping on the hyperfine-split ground states, we propose a simple, all-optically controllable delay for synchronization of heralded narrow-band photons in a quantum network.
\end{abstract}
\begin{document}

\flushbottom
\maketitle

\thispagestyle{empty}

\section*{Introduction}
Forthcoming quantum networks require various building blocks to perform tasks such as logical operations, or error correction on quantum bits, entanglement generation, distillation and distribution, as well as storage of quantum bits (or qubits) \cite{kimble_quantum_2008,divincenzo_quantum_1998}.
The latter two are important for future long distance quantum networks.
In such a network, entangled quantum states can be used to teleport another unknown quantum state \cite{bennett_teleporting_1993,reindl_all-photonic_2018} or for secure quantum key distribution between two distant communicating partners \cite{ekert_quantum_1991}.

Photonic qubits underlie non-zero absorption in communication channels like air or optical fiber, which lead to an exponential decrease of the transmission probability with distance. 
This limitation can be overcome by integrating so-called quantum repeaters into the network \cite{briegel_quantum_1998}.
In the quantum repeater scheme, quantum memories are crucial to store a photonic state as a stationary qubit and to enable a coincident Bell-state measurement with another qubit.

A wide range of approaches exists to implement quantum memories \cite{simon_quantum_2010}, each with own advantages or disadvantages.
Spins in solid state systems, e.g.~the nitrogen nuclear-spin of a nitrogen vacancy (NV) center \cite{fuchs_quantum_2011}
or the electronic spin of a negatively charged silicon vacancy (SiV$^-$) center \cite{sukachev_silicon-vacancy_2017} in diamond, show long coherence times at the order of milliseconds, however minimal operation times of about \SI{100}{\nano\second} drastically limit the communication speed in a quantum network.
A faster scheme based on Raman scattering in bulk diamond, on the other hand, only allows for storage times on the order of picoseconds, which is limited by the optical phonon lifetime at room temperature \cite{england_storage_2015}.
Solid state memories, like the aforementioned as well as the ones based on rare-earth ions in fibers \cite{saglamyurek_multiplexed_2016} or waveguides \cite{zhong_nanophotonic_2017}, suffer from the permanent coupling to the environment and, in most cases, need to be brought to cryogenic temperatures and high magnetic fields for long memory times.
For future operation of quantum networks and adaption to real-world applications, the experimental complexity of the different modules has to be minimized and, thus, such cryogenic systems are to be avoided. 
In contrast, atomic alkali vapors can be the cornerstone of simple, room temperature quantum memories.
Storage and read-out of weak laser pulses at the single photon level has been demonstrated by Raman scattering at the Cesium (Cs) D$_2$ transitions \cite{england_high-fidelity_2012} but it was shown later that four wave mixing generates a significant noise background in atomic vapors \cite{michelberger_interfacing_2015}.
In contrast, electromagnetically induced transparency (EIT) type experiments remain unaffected by four wave mixing \cite{rakher_prospects_2013} and yield an improved signal-to-noise ratio (SNR) compared to Raman type experiments \cite{wolters_simple_2017}.

Interfacing a single-photon source with alkali vapor transitions is a major step to realize a quantum network.
The combination of dissimilar physical systems in a heterogeneous network promises the best performance by exploiting their respective strengths \cite{sangouard_quantum_2011}.
Among all single-photon emitters, epitaxially grown self-assembled semiconductor quantum dots (QDs) stand out as versatile and highly efficient sources of indistinguishable single photons at up to GHz rates \cite{ding_-demand_2016,somaschi_near-optimal_2016,schlehahn_electrically_2016} and also entangled photon pairs \cite{akopian_entangled_2006, trotta_highly_2014,muller_-demand_2014, zhang_high_2015, chen_wavelength-tunable_2016,huber_strain-tunable_2018}.
QDs can be grown in a wide range of wavelengths \cite{buckley_engineered_2012} and are tunable via strain that is transduced from a piezoelectric material into a semiconductor membrane \cite{ding_tuning_2010, rastelli_controlling_2012, yuan_uniaxial_2018}.
First steps have been taken to establish hybrid interfaces between atomic vapors and QD single photons \cite{akopian_hybrid_2011, ulrich_spectroscopy_2014, jahn_artificial_2015, wildmann_atomic_2015, portalupi_simultaneous_2016,huang_electrically-pumped_2017,widmann_faraday_2018,vural_two-photon_2018} as well as entangled photons \cite{trotta_wavelength-tunable_2016}. 
Single photon spectroscopy was performed in Cesium \cite{ulrich_spectroscopy_2014} and Rubidium \cite{jahn_artificial_2015} vapor, a Faraday anomalous dispersion optical filter prepared tailored narrow-band photons from QD resonance fluorescence \cite{portalupi_simultaneous_2016,widmann_faraday_2018}, and a reduced group velocity of the single photon wave packets was demonstrated by tuning the QD emission between the \SI{6.8}{\giga\hertz} and \SI{9.2}{\giga\hertz} hyperfine-split ground-states of the Rubidium D$_2$ \cite{akopian_hybrid_2011,huang_electrically-pumped_2017} and Cesium D$_1$ \cite{wildmann_atomic_2015,vural_two-photon_2018} transitions, respectively.

In order to make the QD single photon emission compatible to an atomic EIT-type memory, it needs to be tuned precisely to an alkali atom excited state transition.
Tuning by changing the temperature or by the quantum confined Stark effect tends to
affect the single photon coherence, causing broader emission lines due to stronger coupling to phononic modes at higher temperature or excessive charges in the semiconductor crystal environment, respectively.
In contrast, the exciton in the QD should ideally be disconnected from external influences, and therefore situated in an environment at static, low temperature with no other charges close-by.

In this work, the QD emission frequency is precisely controlled by piezoelectric-induced strain.
Single photon spectroscopy is performed to characterize the source properties first, in particular the inhomogeneous linewidth.
Next, the dispersion induced delay of single photons in a Cesium vapor cell is considered, while in the following section this is further investigated with respect to the individual frequency components of the single photons.
As a result, a fast tunable optical delay for short storage times up to \SI{1}{\nano\second} is proposed based on these experimental findings.

\section*{Experiment and Results}

\subsection*{Optical excitation and single photon collection}
\begin{figure}[ht!]
\centering\includegraphics[width=1.0\textwidth]{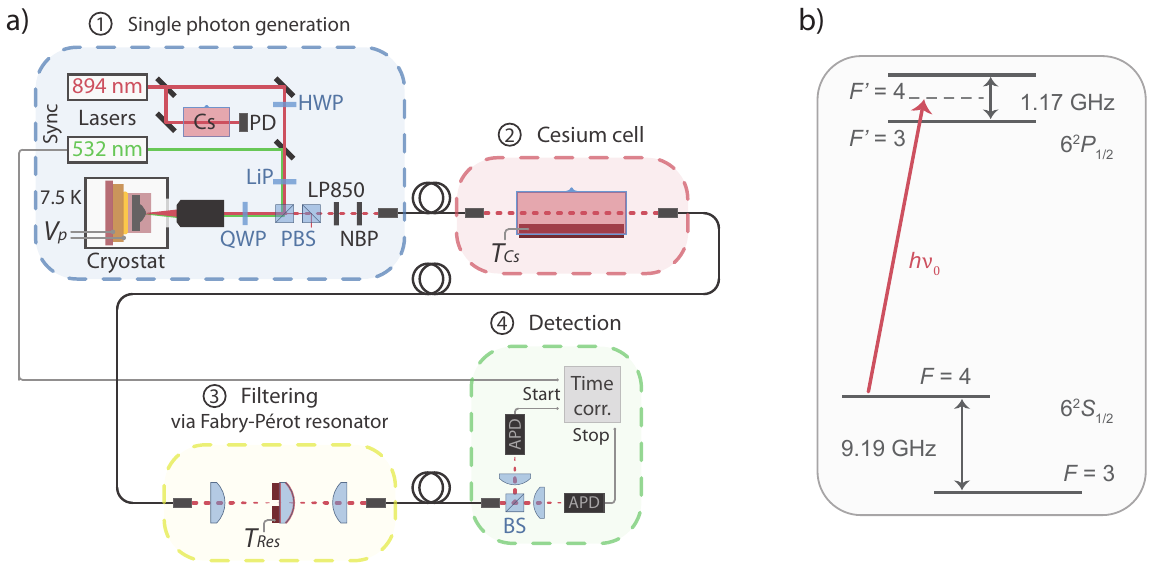}
\caption{(a) Experimental setup. Single photons are generated by resonant (\SI{894}{\nano\metre}) or non-resonant (\SI{532}{\nano\metre}) excitation of an In(Ga)As QD in a liquid-Helium flow cryostat (1). The emission frequency of the QD exciton is strain-tuned via the piezo voltage $V_p$. The resonant laser can be scanned over the four hyperfine-split Cs D$_1$ transitions. The pump laser is strongly suppressed by polarization optics: half-wave plate (HWP), linear polarizer (LiP), polarizing beam splitters (PBS), and quarter-wave plate (QWP). In case of non-resonant excitation, the single photons are further filtered from pump light and photons from other QD states with an \SI{850}{\nano\metre} longpass (LP850) and a \SI{1}{\nano\metre} narrow bandpass (NBP) at \SI{894}{\nano\metre}, before being coupled into a fiber. From there, the QD single photons can be sent to different sections -- a temperature controlled, shielded Cs cell (2), a monolithic Fabry-P\'erot resonator for spectral filtering (3), and the detection part (4), consisting of a 50/50 beamsplitter (BS), two avalanche photo diodes (APD), and time-correlation electronics. (b) Energy levels of the Cs D$_1$ line. For most of the experiments the QD emission frequency was centered to $\nu_0$ between the $F=4\to F^\prime=3$ and $F=4\to F^\prime=4$ transitions.}
\label{fig:setup}
\end{figure}

Figure~\ref{fig:setup} introduces the complete experimental setup.
The QD sample is cooled to \SI{7.5}{\kelvin} in a liquid-helium flow-cryostat (Cryovac Konti Micro) with optical access (part (1) in Fig.~\ref{fig:setup} (a)).
For excitation and detection of single QDs a high NA microscope objective (LMPlanIR, $N\! A=\SI{0.8}{}$, Olympus) is used.
For resonant excitation of the QD, the \SI{894}{\nano\metre} laser light (EYP-DFB-0894, eagleyard Photonics) is suppressed by a cross-polarization configuration prior to detection (see A. Kuhlmann \textit{et al.}~\cite{kuhlmann_dark-field_2013}).

By using motorized rotation mounts with an angular accuracy of $10^{-3}$~degrees (Newport, Conex AG-PR100P) to adjust the linear polarizer and the quarter-wave plate (QWP), we achieve an overall suppression of reflected laser intensity in resonance fluorescence experiments at the order of $10^{6}$, before coupling to a single mode fiber.
This way, signal to noise ratios (SNRs) between the photons scattered at the QD and unfiltered laser background of up to $SNR=30$ were achieved.
A pulsed \SI{532}{\nano\metre} laser (PicoQuant LDH-P-FA-530, PDL 800-D) is either used at power below \SI{1}{\nano\watt} for repumping the QD in case of charge carrier trapping under resonant excitation, or with a \SI{40}{\mega\hertz} repetition rate for pulsed non-resonant excitation of the QD.

\subsection*{Single photon spectroscopy on the Cesium D$_1$ line}
\label{char-spectro}
\begin{figure}[ht!]
\begin{center}
\includegraphics[width=1.00\textwidth]{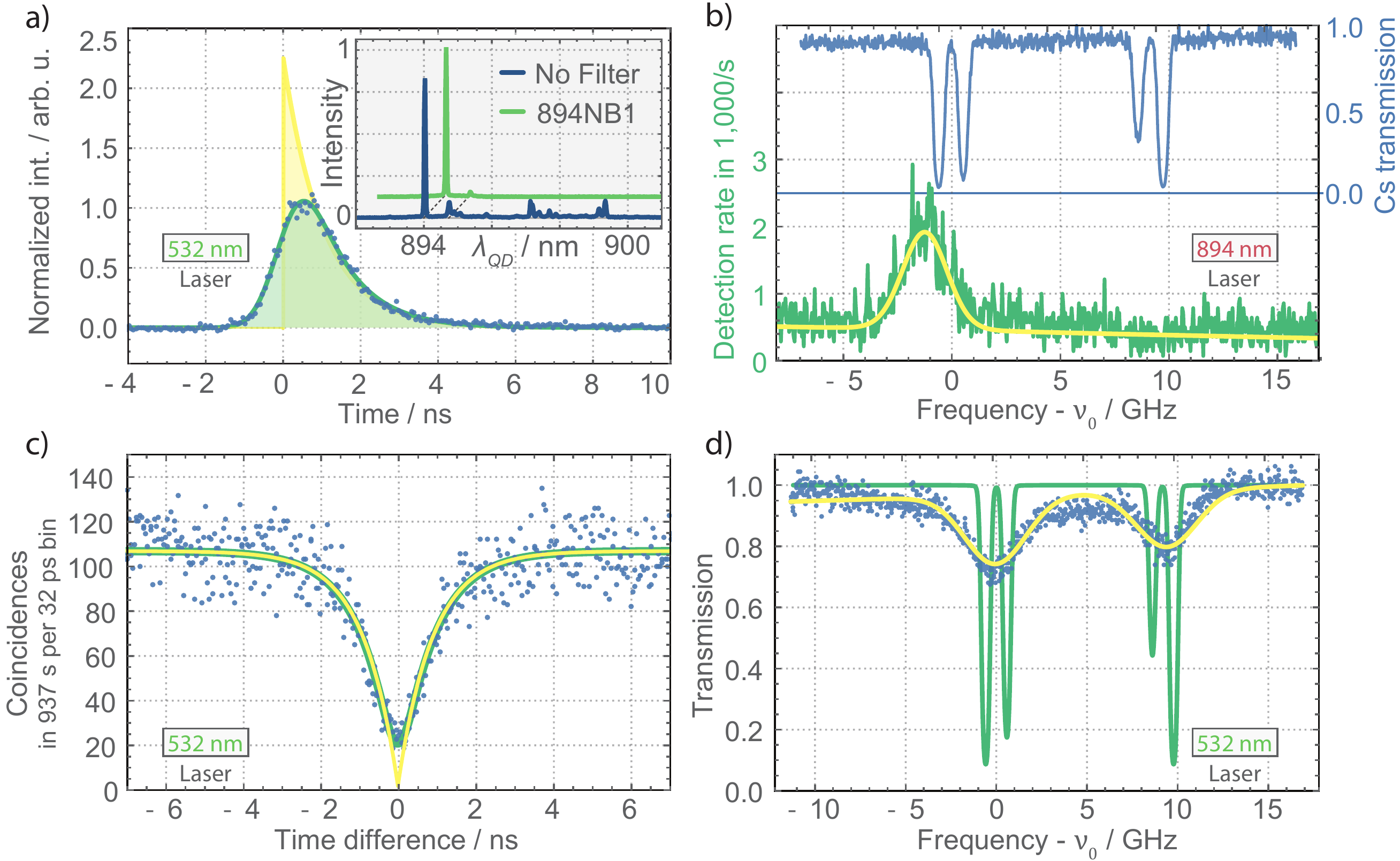}
\end{center}
\caption{Overview of the QD properties. (a) Non-resonantly excited lifetime measurement of the QD exciton. A convolution of an exponential decay and the Gaussian timing instrument response (green) is fit to the data (blue). The resulting decon- voluted decay $\propto \exp{(-t/\SI{1.04}{\nano\second})}$ is shown in yellow. Inset: QD spectrum with (green) and without (blue) \SI{1}{\nano\metre} bandpass filter at \SI{894.335}{\nano\metre}. (b) Independent scans of the \SI{894}{\nano\metre} laser over the QD resonance (green) and the Cs D$_1$ transitions (blue). A Voigt fit (yellow) to the QD exciton spectrum reveals a FWHM linewidth of $\SI{2.4\pm0.2}{\giga\hertz}$. (c) The QD exciton emission under non-resonant excitation shows strong anti-bunching (blue data) of $g^{(2)}_{\textrm{data}}(0)=0.19\pm0.03$ (green), corresponding to $g^{(2)}_{\textrm{deconv}}(0)=0.02\pm {}^{0.03} _{0.02}$ after deconvolution of the timing jitter (yellow). (d) Absorption of the non-resonantly excited QD photons in Cs vapor at $\vartheta_{Cs}=\SI{30}{\celsius}$. The QD line was scanned over the Cs D$_1$ spectrum. Fitting a convolution of the simulated Cs D$_1$ lines (green) and a Voigtian QD spectrum to the transmission data (blue) yields a QD linewidth of $\SI{3.6\pm.1}{\giga\hertz}$ under non-resonant excitation, which has an additional Gaussian broadening of about $\SI{2.7\pm.1}{\giga\hertz}$ FWHM compared to the resonant scan in (b).}
\label{fig:qdoverview}
\end{figure}

First a single QD was characterized by performing single photon spectroscopy on the Cesium D$_1$ line.
The excitonic QD emission was tuned to the Cesium D$_1$ transition at \SI{894.335}{\nano\metre} (see Methods).
The QD spectrum is shown in the inset in Fig.~\ref{fig:qdoverview}~(a) under non-resonant excitation well below saturation.
The exciton line was filtered with a \SI{1}{\nano\metre} bandpass filter for the following experiments.

To measure the QD linewidth under resonant excitation, the \SI{894}{\nano\metre} laser is scanned over a frequency interval of \SI{25}{\giga\hertz} around the Cesium D$_1$ line.
It generates resonantly scattered single photons from the QD exciton (green spectrum in Fig.~\ref{fig:qdoverview}~(b)).
The Cesium transmission spectrum (blue) was measured for reference on a photo diode by sending the scanning laser through a Cesium cell at \SI{35}{\celsius}. 
The Voigt fitting function (yellow) consists of a Lorentzian with a fixed homogeneous linewidth of $\Delta\nu_{hom}= 1/(2\pi T_1)=\SI{153}{\mega\hertz}$ (deduced from the lifetime measurement in Fig.~\ref{fig:qdoverview}~(a)) and the Gaussian width as a free parameter to take inhomogeneous broadening from spectral diffusion into account.
A linear part is added to the fit function for slightly frequency-dependent changes in the laser suppression.
The extracted value of the inhomogeneous broadening under resonant excitation is $\Delta\nu_{inhom - res}=2.4\pm \SI{0.2}{\giga\hertz}$.

The autocorrelation function of the exciton emission under non-resonant pumping was measured by correlating the photon detections on the two single-photon counting modules (SPCMs) based on Si avalanche photo diodes (APDs) in Hanbury Brown and Twiss configuration.
The single photon count rates were $R_{APD1}=76000 \; 1/\textrm{s}$ and $R_{APD2}=46000\; 1/\textrm{s}$.
Coincidences are plotted in a histogram by their time-of-arrival differences (blue dots in Fig.~\ref{fig:qdoverview}~(c)). 
The data was fitted with the convolution (green curve) of an exponential dip, typical for QD single photons\cite{zwiller_single_2001,moreau_quantum_2001, yuan_electrically_2002,buckley_engineered_2012,kantner_hybrid_2017}, and the instrument response function (IRF, see Methods for details).
The dip around zero time difference indicates strong antibunching of $g^{(2)}_{\textrm{data}}(0)=0.19\pm0.03 <0.5$ and proves the dominant single photon character of the light, as directly measured.
The deconvolution of the IRF yields an even lower value of $g^{(2)}_{\textrm{deconv}}(0)=0.02\pm {}^{0.03} _{0.02}$, close to perfect single photons (yellow).

Spectroscopy at a very well-known and well-described system allows for the detailed investigation of the involved photons.
Single photon spectroscopy at the Cesium D$_1$ lines was performed with QD photons under continuous wave, non-resonant pumping.
The light from the QD was collected into a fiber and sent through the Cesium cell at a temperature of $\vartheta_{Cs}=\SI{30}{\celsius}$ (part (2) in Fig.~\ref{fig:setup}).
The QD emission frequency was tuned over the Cesium spectrum at constant speed, while the single photon count rate was recorded.
The transmission of the QD photons is reduced by absorption at the Cesium D$_1$ transitions (blue data points in Fig.~\ref{fig:qdoverview}~(d)). 
For reference, the calculated (ElecSus software \cite{zentile_elecsus:_2015,zentile_elecsus:_2015-2}) transmission spectrum of the Cesium cell at \SI{30}{\celsius} is shown (green line).
The data is fitted with a convolution of the atomic transmission spectrum with the QD emission (yellow in Fig.~\ref{fig:qdoverview}~(d)).
Here, an additional linear term is added to account for a second QD emission line at \SI{895}{\nano\metre}.
This emission is visible as a small peak in the spectrum (Fig.~\ref{fig:qdoverview}~(a) inset) and partially transmitted by the narrowband filter, in particular when higher piezo voltages are applied.

The QD emission spectrum is modeled by a Voigt profile consisting of a Lorentzian function with the QD natural linewidth and a Gaussian linewidth as a free fitting parameter.
Under non-resonant excitation the QD emission has a lifetime of $T_1=1.04\pm\SI{0.02}{\nano\second}$ (Fig.~\ref{fig:qdoverview}~(a)), which corresponds to a natural homogeneous linewidth of $\Delta\nu_{hom}= 1/(2\pi T_1)=153\pm\SI{3}{\mega\hertz}$, assuming no additional dephasing of the QD exciton state. 
Under non-resonant excitation a large amount of excess electrons and holes are generated in the vicinity of the QD, that cause inhomogeneous line broadening due to spectral diffusion based on the Coulomb interaction between the QD exciton and the surrounding charges \cite{liu_optical_2017}.
Taking spectral diffusion into account, the total inhomogeneous broadening is determined to be $\Delta\nu_{inhom-non-res}=\SI{3.57\pm.08}{\giga\hertz}$ under non-resonant excitation, which dominates the overall QD FWHM linewidth of $\Delta\nu_{QD}=\SI{3.65\pm.16}{\giga\hertz}$.
This corresponds to an additional Gaussian broadening of $\Delta\nu_{non-res}=\SI{2.7\pm.1}{\giga\hertz}$ compared to resonant excitation (Fig.~\ref{fig:qdoverview}~(b)).

\subsection*{Single photon delay between the excited state Cesium D$_1$ transitions}
\label{delay-temp}
\begin{figure}[ht!]
\begin{center}
\includegraphics[width=1.00\textwidth]{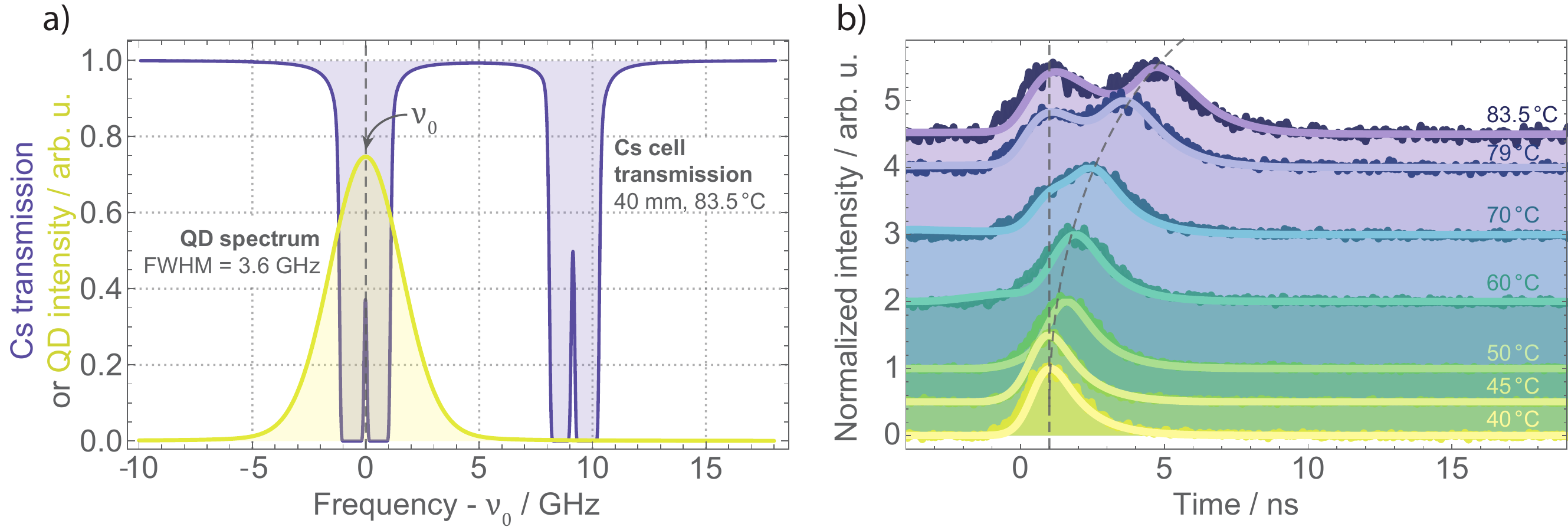}
\end{center}
\caption{Delay of single photons between the hyperfine-split excited Cs D$_1$ transitions. (a) The QD exciton emission spectrum (yellow) is tuned in the middle of the $F=4$ ground state and both excited states (purple). (b) With rising temperature of the Cs vapor, the group velocity of the single photon is gradually decreased. This is demonstrated in the experiment by later detection events with increasing temperature. Simulations of the pulse transfer in the Cs cell (see Methods) are in excellent agreement with the measurements. The dashed lines are guides to the eye.}
\label{fig:nonresdelay}
\end{figure}

The central frequency of the QD exciton emission is tuned to the middle of the two excited state transitions $6^2\text{S}_{1/2} F=4 \to 6^2\text{P}_{1/2} F^\prime=3$ and $6^2\text{S}_{1/2} F=4 \to 6^2\text{P}_{1/2} F^\prime=4$ of Cesium, as indicated by $\nu_0$ in Fig.~\ref{fig:nonresdelay}~(a). 
Note that in this case a large part of the QD spectrum now lies between the two transitions in a region with low absorption.
At this central frequency, the QD photons are sent through the Cesium cell and detected (parts (2) and (4) in Fig.~\ref{fig:setup}, respectively).
The photon arrival times are correlated and histogrammed (Fig.~\ref{fig:nonresdelay}~(b)) with respect to a preceding trigger signal from the pulsed \SI{532}{\nano\metre} laser.
The temperature of the Cesium vapor was varied between 35 and \SI{83.5}{\celsius}.

For low Cesium temperatures (\SI{40}{\celsius} and \SI{45}{\celsius}), at low optical densities of the vapor, the measurements resemble the reference measurement with the Cesium cell removed (Fig.~\ref{fig:qdoverview}~(a)). 
With increasing temperature a feature is emerging, which represents the delayed fraction of the single photon wave packet.
The solid lines superimposed on the data are simulation results, calculated by pulse propagation of the \SI{3.6}{\giga\hertz} wide QD spectrum through the Cesium cell (see Methods section).
The Cesium vapor temperature defines the retardation time of the delayed components of the photon.
The only free parameters for fitting the simulation to the data are the normalized pulse amplitude, the background, and slight variations of the QD central frequency $\ll\SI{1}{\giga\hertz}$ to adjust for the ratio of peak heights of the delayed and non-delayed pulse components.
With these tiny adjustments, the theoretical simulations perfectly match the collected data. 

Even at the highest temperatures, corresponding to a peak optical density of OD~$=69$ on the $F=4\to F^\prime=3$ transition, there is still a window of \SI{35}{\percent} transmission (or OD~$=0.43$) at frequency $\nu_0$ with a bandwidth of about $\SI{150}{\mega\hertz}$ between the two absorption lines (Fig.~\ref{fig:nonresdelay}~(a)). 
This allows for delay of Fourier-limited QD photons under optical control, as discussed in the last section ``Optically controlled delay for single photons''.

From here on, we will treat the delay time $\langle \tau_{d} \rangle$ of a photon propagating through the Cesium cell as the retardation of the average detection time behind the cell $\langle \tau_{cell} \rangle$ with respect to a reference photon $\langle \tau_{ref} \rangle$ transmitted through air: 
$\langle \tau_{d} \rangle=\langle \tau_{cell} \rangle-\langle \tau_{ref} \rangle$ .
In the experiment and simulation we define the average detection time $\langle \tau_{i} \rangle$ of a photon as the center of mass of its temporal wave packet.

\subsection*{Spectrally resolved single photon delay}
\label{delay-spec}
To investigate the spectral dependency of the single photon delay, the QD single photons are sent through the Cesium cell, as before, and subsequently filtered by the Fabry-P\'erot resonator (cf. Fig.~\ref{fig:setup} part (3)).
The resonator has a linewidth of $\Delta\nu_{FP}=\SI{192}{\mega\hertz}$ (cf. Methods section) which is one order of magnitude smaller than the QD's inhomogeneous exciton linewidth of \SI{3.6}{\giga\hertz} and very close to the Fourier-limited width of \SI{153}{\mega\hertz}.
This allows for resolving the atom induced delay of the individual frequency components.
\begin{figure}[ht!]
\centering\includegraphics[width=1.0\textwidth]{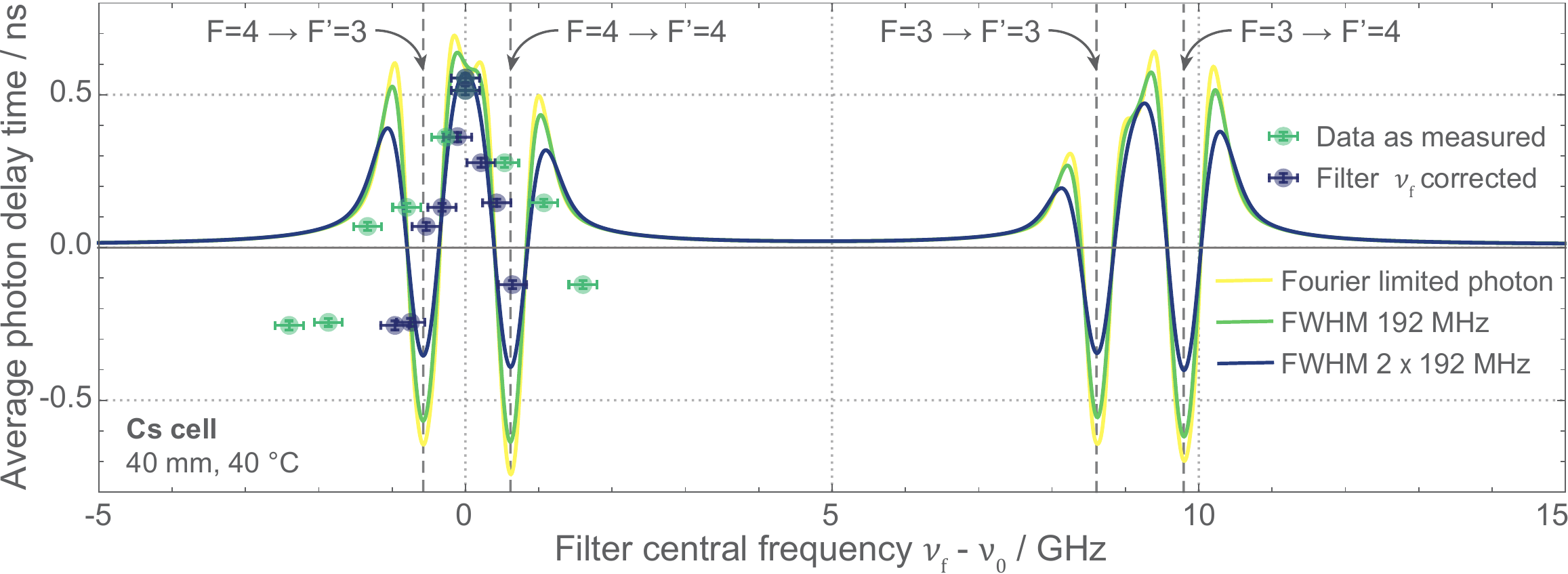}
\caption{Spectrally resolved delay of a single photon. The photon propagates through the \SI{40}{\milli\metre} Cs vapor cell at \SI{40}{\celsius}. 
Only the frequency components of the QD spectrum are detected which pass the Fabry-P\'erot filter around its transmission frequency~$\nu_f$. The average value for the detection time $\langle \tau_{d} \rangle$ of a photon wave packet (green data points) is displayed as the difference to a photon traversing \SI{40}{\milli\metre} of air. See text for details on corrected filter frequency $\nu_f$ (blue data points). Frequency components around $\nu_f=\nu_0$ of the QD spectrum are delayed, while others near the resonances (dashed vertical lines) at about $\pm\SI{0.6}{\giga\hertz}$ appear as ``fast light'' due to pulse distortion. Theory is calculated for Fourier-limited photons \quad (yellow curve), photons of the filter transmission width (green curve), and photons with \SI{384}{\mega\hertz} FWHM (blue curve).}
\label{fig:filteringdelay}
\end{figure}

The QD exciton is tuned to $\nu_0$ (Fig.~\ref{fig:nonresdelay} (a)) and the Cesium vapor temperature is set to $\vartheta_{Cs}=\SI{40}{\celsius}$.
The \SI{3.6}{\giga\hertz} broad exciton line spans across both nearest transitions.
One transmission line of the Fabry-P\'erot resonator is set to $\nu_f=\nu_0$. 
In the experiment different spectral components are addressed by tuning the transmission frequency of the filter resonance via the resonator's temperature.
After the first measurement at $\nu_f=\nu_0$ the resonator temperature was successively decreased and data was acquired at higher frequencies $\nu_f$ (green data points in Fig.~\ref{fig:filteringdelay}). 
The resonator was then set back to $\nu_f=\nu_0$ and measurements at subsequent, lower frequencies (higher resonator temperatures) were taken.
As a result of the progression of the experiment, the resonator did not fully thermalize, leading to an overestimation of the frequency offset $| \nu_f-\nu_0 |$. 
We compensate for this effect by using a reduced resonator tuning coefficient $\tilde \delta = 0.4 \, \delta$.
With this correction, the blue data points are in good agreement with theoretical expectations (solid lines), in particular when assuming that the resonator linewidth is effectively increased due to the ongoing thermalization (blue curve).

Propagation of a pulse through a medium with pronounced dispersion and absorption leads to considerable distortion of the wave packet. 
The time of the amplitude maximum of a delayed and distorted wave packet does not describe its temporal behavior well enough anymore.
Instead, we consider its center of mass to represent the average detection time of the photon.
In the following analysis we derive the ``delay time'' $\langle \tau_{d} \rangle=\langle \tau_{cell} \rangle-\langle \tau_{ref} \rangle$ from the center-of-mass difference of a delayed and a reference photon, which is calculated as the weighted average of the temporal wave packets, for both the experimental histogram data and the simulated wave forms.
However, this reduction to a single parameter, i.e. pulse delay time, may not be appropriate and can lead to unphysical interpretations. 

The average photon delay $\langle \tau_{d} \rangle$ at frequency $\nu_f=\nu_0$ is greater than that of a photon traveling the same distance in air.
This is well described by the reduced group velocity of the photon wave packet due to the strong dispersion between nearby resonances.
In the regions of the two Cesium resonances at about $\nu_f-\nu_0 = \pm \SI{0.6}{\giga\hertz}$ negative average delays indicate the well-known regime of ``fast light'' \cite{dogariu_transparent_2001,stenner_speed_2003,milonni_fast_2004}. 
Upon measurement of the average arrival time, which was broadened by the IRF, this creates the false impression that the photon traveled faster than light in vacuum. 
In such a regime, as pointed out above, the description of pulse distortion by a single parameter ``delay time'' is meaningless. 
Instead experimental data has to be compared to careful numerical simulations (e.g. Fig.~\ref{fig:nonresdelay} (b)).
The simulated transferred pulses confirm that the front velocity, as discussed in Ref.~\cite{milonni_fast_2004}, is always the vacuum speed of light.
It also demonstrates that the ``fast light'' behaviour can be understood as the early components of the photon wave packet being transmitted, while later components are absorbed. 
Only the delay of light at a frequency $\nu_f=\nu_0$ in the transmission window between the two resonances is mainly due to dispersion.

The main distortion of the single photon wave packet in delay experiments in warm atomic vapor is caused by absorption of particular frequencies by the sidebands of nearby resonances.
For the use of these delays in optical quantum networks it imposes the condition that the bandwidths of the individual photons should lie well outside of the absorptive features which provide the dispersion.
Otherwise the wave packet of a delayed photon would be strongly distorted, making it clearly distinguishable from an undelayed one. 
Obviously, this renders subsequent two-photon interference, an often needed operation in quantum networks, impossible. 

\subsection*{Optically controlled delay for single photons}
\label{delay-switch}
Changing the temperature is not the only way to control the optical density and propagation delay in the vapor cell.
Alternatively, for the demonstrated delay between two excited states, it is possible to modify the OD by changing the ground state population via optical pumping.
\begin{figure}[ht!]
\centering\includegraphics[width=1.0\textwidth]{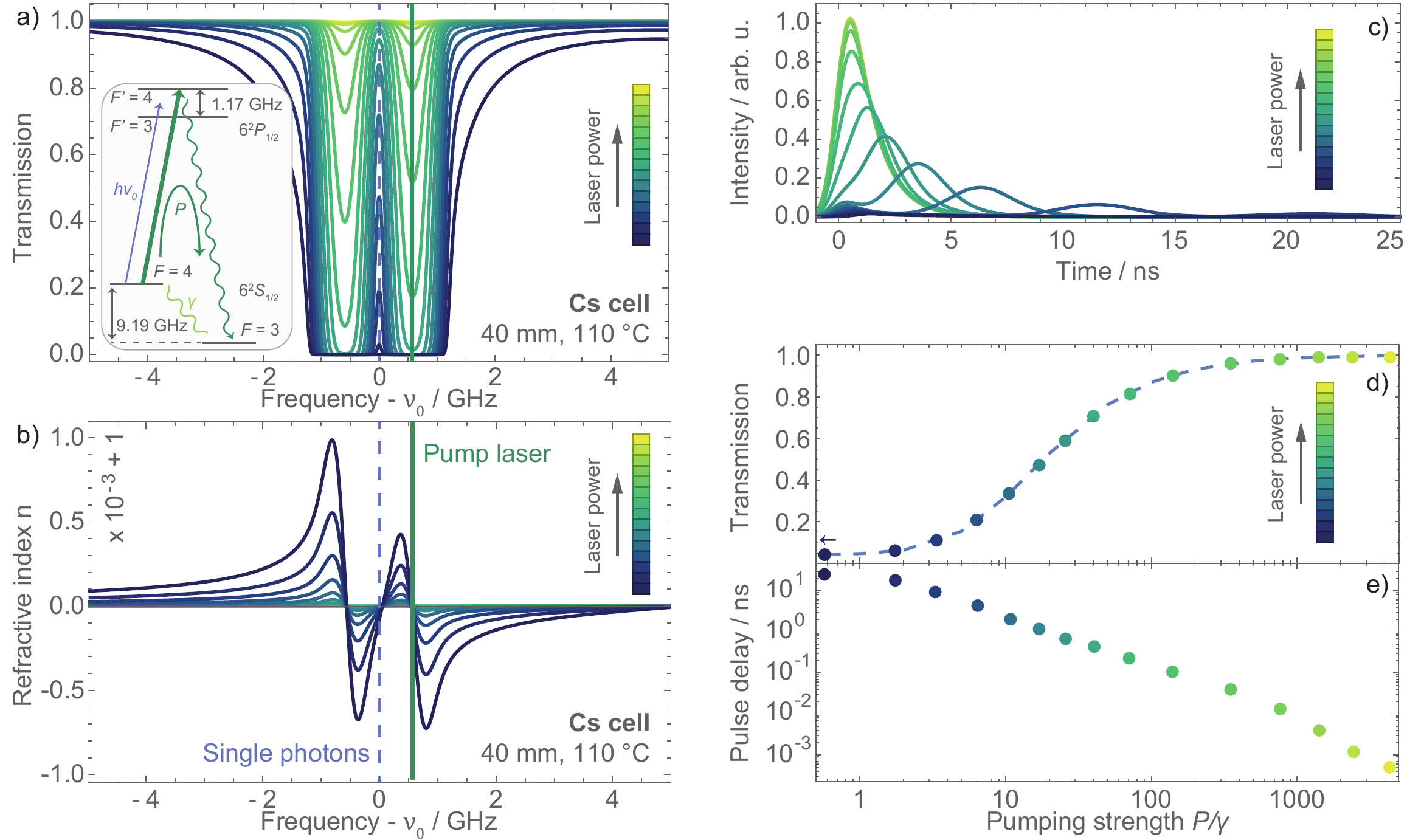}
\caption{Simulation for optical control of the single photon delay. (a) The power of a pump laser resonant with the $F=4\to F^\prime=4$ transition (green in inset) controls the effective optical density of the atomic vapor for a single photon of frequency $\nu_0$ via the population of the $F=3$ and $F=4$ ground states. The gradient of the refractive index (b) and the temporal distribution of a delayed single photon (c) change accordingly. For high laser powers the total transmission (d) of a \SI{192}{\mega\hertz} wide photon approaches 1, but the delay achieved is becoming increasingly small (e).}
\label{fig:delay-switch}
\end{figure}

In a Cesium cell without external fields, the two hyperfine-split ground state populations $\rho_{F=3}\equiv\rho_{33}$ and $\rho_{F=4}\equiv\rho_{44}$, in the notation of a quasi two-level system density matrix, approximately thermalize to $\tilde\rho_{33}\approx\tilde\rho_{44}\approx0.5$ with the depolarization rate $\gamma$.
In room temperature vapor cells this is mainly due to spin flips caused by atom-atom collisions or atom-wall collisions.
If now a pump laser is exciting one of the two transitions on resonance (e.g. $F=4 \to F^\prime=4$ in Fig.~\ref{fig:delay-switch} (a) inset), it will transfer population density from $F=4$ to the other ground state $F=3$ with an effective pump rate $P$.
The pump rate is assumed to be proportional to the laser intensity, $P=\beta I$, via the pumping coefficient $\beta$.
The population of both states in dependence of the pump laser intensity is determined by the solution of the temporal differential equations: 
\begin{align}
    \dot \rho_{33}&=+\gamma \rho_{44}-\gamma \rho_{33}+P\rho_{44}\\
    \dot \rho_{44}&=-\gamma \rho_{44}+\gamma \rho_{33}-P\rho_{44} \; .
\end{align}
After thermalization, the equilibrium solutions $\bar\rho_{33}$ and $\bar\rho_{44}$ depend only on the pumping strength with respect to the depolarization rate $P/\gamma$.

The calculations presented in Fig.~\ref{fig:delay-switch} were made for the same Cesium cell as in the previous experiments, but with a temperature of $\vartheta_{Cs}=\SI{110}{\celsius}$.
The reduction of the ground state population $\bar\rho_{44}$ reduces the optical density and increases the transmission through the Cesium cell (Fig.~\ref{fig:delay-switch} (a)) according to $\exp{(-L \cdot\alpha(\nu)\cdot \bar\rho_{44}/\tilde\rho_{44})}$.
Here, $L$ is the length of the Cesium cell and the absorption coefficient $\alpha(\nu)$ is deduced from the electric susceptibility.
The respective effect on the refractive index (Fig.~\ref{fig:delay-switch} (b)), and therefore the dispersion of the single photon in the atomic vapor, is mimicked by calculating the susceptibility for lower vapor temperatures, which correspond to the reduced optical densities.
The OD  as well as the dispersion between the two considered resonances are strongly reduced with increasing pumping strength $P/\gamma$.

The incoming single photon pulse is again defined by the excitonic lifetime of \SI{1.04}{\nano\second}, its central frequency $\nu_0$, and inhomogeneous broadening to a spectral width of \SI{3.6}{\giga\hertz}.
In the simulation, the pulse is filtered down to \SI{192}{\mega\hertz} bandwidth by a Fabry-P\'erot resonator.
After transmission through the Cesium cell, detection with the same timing jitter of about \SI{1}{\nano\second} is assumed as before.
The resulting temporal shapes of the transmitted photon wave packet are displayed in Fig.~\ref{fig:delay-switch} (c).
For each pumping strength the overall transmission (Fig.~\ref{fig:delay-switch} (d)) is calculated as the integral over the full wave packet after leaving the cell, normalized by the input pulse area.
The pulse delay (Fig.~\ref{fig:delay-switch} (e)) is expressed by the retardation of the average photon detection time with respect to the input pulse.

Under the above-mentioned conditions and a pumping rate $P$ at the same order as the depolarization rate $\gamma$
, the average single photon detection time is delayed for about \SI{25}{\nano\second} with some components being delayed for up to \SI{80}{\nano\second}.
But almost no light is transmitted through the Cesium cell.
With increasing pump laser power (lightening colors in Fig.~\ref{fig:delay-switch}) the ground state population is gradually transferred to the $F=3$ hyperfine state and the optical density for the single photon between the resonances $F=4\to F^\prime =3$ and $F=4\to F^\prime =4$ converges to zero. 
The pulse delay is correspondingly reduced and the transmission of the photon goes to one.
The temporal shape of the photon, however, elongates with larger delay times because of the narrowing of the transmission window between the two resonances and the associated additional spectral filtering of the photon.
The temporal width of the main peak at about $\SI{6}{\nano\second}$ in the delayed light (blue in Fig.~\ref{fig:delay-switch} (c)), for instance, is already twice as wide as the undelayed pulse (yellow) at a pumping strength of $P/\gamma\approx 15$, where the total transmission is about $50~\%$. 
Additionally, a part of the wave packet remains close to zero temporal delay, which corresponds to frequencies in the broad shoulders of the Lorentzian Fabry-P\'erot filter.
Altogether, the dispersive delay of light drastically changes the temporal shape of photons if portions of their spectra are absorbed by nearby resonances.

This configuration could be used as an optically switchable delay in the region where only small losses, e.g.~$<\SI{3}{\deci\bel}$, reduce the probability of detecting the delayed photon and if the spectrum of the utilized light source is narrower than the dispersive window.
Switching between different delay times could be performed at timescales of about \SI{100}{\nano\second} just above the lifetime of the excited state $6^2P_{1/2}$ $(T\approx \SI{35}{\nano\second})$.
Delay times at the order of one nanosecond allow for synchronization in small scale quantum networks.

Such a switchable delay is especially interesting for applications with narrow-band photon pair sources based on spontaneous parametric down-conversion~\cite{ahlrichs_bright_2016}.
The random process of photon pair generation can be pushed towards deterministic emission of one photon.
In such a scheme one photon of a pair is detected and heralding the other, so called signal photon.
The switchable delay therefore enables the deterministic emission of an heralded signal photon synchronously to other processes of the quantum network \cite{nunn_enhancing_2013}.

\section*{Conclusion}

To summarize, we have tuned the single photon emission from an In(Ga)As QD exciton precisely to the hyperfine-split excited state transitions of the Cesium D$_1$ line.
The applied fine tuning mechanism by piezoelectric-induced strain of the semiconductor lattice demonstrates that QDs are potent photon sources, capable to be interfaced with atom based quantum nodes with an accuracy at the order of the atomic linewidth.
Single photon spectroscopy was performed at the well-known D$_1$ transitions in Cesium to characterize the spectral properties of the QD emission under resonant and non-resonant excitation.

The QD single photon was delayed by up to \SI{5}{\nano\second} between the $F=4\to F^\prime=3$ and $F=4\to F^\prime=4$ transitions by dispersive reduction of the group velocity.
Line broadening of the QD spectrum by spectral diffusion limits the delay efficiency here.
The single photon was only delayed when the exciton energy lay in the region of low group velocity between the two resonances.
Spectrally resolved delay measurements, utilizing a monolithic Fabry-P\'erot resonator with transmission width almost identical with the Fourier-limited QD linewidth, provide insight into the interaction of different frequency components from the QD emission with the atomic vapor.
Off-center photons from the QD spectrum with respect to the two atomic resonances experience a strong reduction of the dispersive delay down to zero.
This is due to absorption of specific spectral components, which results in extraction of probability amplitude from the temporal single photon wave packet.
In the most extreme case this led to the detection of ``fast light''.

The reliability and quality of single-photon sources based on QDs has increased significantly in the last years. 
Bright sources providing Fourier-limited photons can be perfectly matched to transitions in alkali vapor cells. 
On the one hand this provides convenient room-temperature quantum memories \cite{wolters_simple_2017}. 
On the other hand the atomic resonances and optical pumping of atomic populations introduce a way to modify the dispersive and absorptive behavior and thus introduce a versatile element for propagation control of single photons in future quantum networks.  

\section*{Methods}
\subsection*{Quantum dot sample}
We use MBE grown In(Ga)As QDs embedded in the middle of a \SI{300}{\nano\metre} thick GaAs membrane.
The membrane is attached to a PMN-PT piezo-electric actuator.
See Ref.~\cite{rastelli_controlling_2012} for details.
The QD emission wavelength can be fine-tuned by strain-induced change of the electric band-structure of the semiconductor material.
The piezo-electric element is oriented in a way that, under application of voltage, it induces biaxial stress to the QDs
and allows for a frequency-tuning of the excitonic emission lines. 
The excitonic emission lines of QDs on this sample lie within a range of $\SI{895\pm10}{\nano\metre}$ and have a tuning coefficient of about \SI{625}{\mega\hertz\per\volt} allowing for tuning by up to \SI{1.5}{\nano\metre} for a maximum voltage range of \SIrange{-300}{600}{\volt} .

\subsection*{Cesium cell}
In this experiment we use 99.99$\%$ isotopically pure $^{133}$Cs \cite{hockel_electromagnetically_2010} in a \SI{4}{\centi\metre} long quartz glass cell with a diameter of \SI{25}{\milli\metre} (part (2) in Fig.~\ref{fig:setup}), with anti-reflection (AR) coating at the front and end facet.
The optical density (OD) of the Cesium can be tuned by adjusting the temperature in a range of $\vartheta_{Cs}=$ \SIrange{25}{80}{\celsius} with a temperature controller (Meerstetter, TEC-1091).
The Cesium cell and heating foil are placed inside three layers of $\mu$-metal, which reduce the Zeeman shift due to earth's magnetic field on the energies of the Cesium ground and excited states by roughly five orders of magnitude \cite{sterne_multilamellar_1935}, down to a few Hz.

\subsection*{Narrow-band spectral filtering with a monolithic resonator}
\label{sec:filter}
The spectral dependency of the photon delay in Cesium vapor is investigated with a high-resolution optical filter.
In this work we use a monolithic Fabry-P\'erot resonator \cite{ahlrichs_monolithic_2013} with precise tunability and excellent long-time stability. 
It has a free spectral range of $\text{FSR}=\SI{37.8}{\giga\hertz}$, a cavity linewidth of $\Delta\nu_{FP}=\SI{192}{\mega\hertz}$, and a resulting finesse $\mathcal{F}\approx200$.
By careful modematching, a transmission through the resonator of $T\approx \SI{50}{\percent}$ is achieved.
Details on the resonator can be found in Ref.~\cite{ahlrichs_monolithic_2013}.

For tuning the filter frequency $\nu_f=\nu_0+\delta\cdot\Delta\vartheta_f$, the resonator temperature $\vartheta_f$ is adjusted by a temperature controller. 
The resonator frequency was calibrated with a \SI{894}{nm} test laser (Toptica, TA Pro), locked to the four different hyperfine transitions of the Cesium D$_1$ line.
We measured a linear tuning coefficient of $\delta=\SI{-2.64 \pm 0.02}{\giga\hertz\per\kelvin}$ in the range of the Cesium D$_1$ transitions.

\subsection*{Detection}
\label{sec:detect}
The QD emission is detected by two avalanche photodiodes (APD, Excelitas, SPCM AQRH 14) at the two output ports of a 50/50 beamsplitter (BS, part (4) in Fig.~\ref{fig:setup}). 
The electrical TTL pulses are counted by time correlating electronics (Picoquant, PicoHarp 300).
The time correlator measures the time difference between Start and Stop input and generates a histogram from this data.

Depending on the experiment, either the APD~1 output or an electrical synchronization signal from the pulsed \SI{532}{\nano\metre} laser is used to start the time measurement, while APD~2 is always connected to the Stop channel.
In the first case, the autocorrelation statistics of the emitter is measured in the Hanbury Brown and Twiss configuration \cite{walls_evidence_1979}.
The value of the normalized autocorrelation function at zero time difference between two detections, $g^{(2)}(\Delta\tau=0)$, determines the single photon characteristics of an emitter.
For an ideal single photon emitter, without background photons from the environment and dark counts, it would yield $g^{(2)}(0)=0$.
The temporal instrument response function (IRF) of the complete detection system is a Gaussian with a FWHM temporal jitter of $\tau_{IRF}=\SI{1060}{\pico\second}$.
Alternatively to detection by APDs, a Princeton Instruments Acton SP2500 spectrometer with Andor iDus camera is used for measurement of spectra.

\subsection*{Simulation}
The transmission of a wave packet $\mathbf{E}_{in}(\nu)$ through an optical medium can be calculated as $\mathbf{E}_{out}(\nu)=T(\nu)\cdot \mathbf{E}_{in}(\nu)$ with the transfer function $T(\nu)=\exp(i\, n_c\, k\, L)$, including the complex refractive index $n_c$, the wavenumber $k$, and the length of the optical medium $L$.
The complex refractive index $n_c(\nu)=n(\nu)+i/2k\cdot\alpha(\nu)=\sqrt{1+\chi(\nu)}$ is calculated from the electric susceptibility $\chi(\nu)$ for Cesium vapor under experimental conditions \cite{zentile_elecsus:_2015,zentile_elecsus:_2015-2}.
The wave packet will have a group velocity that is frequency dependent via the refractive index: $v_g(\omega)=c_0/(n(\omega)+\omega \frac{\partial n(\omega)}{\partial \omega})$.

For a QD single photon from recombination of an exciton with the lifetime $T_1$ a temporal wave packet $\mathbf{E}_{in}(t)=\mathbf{E}_0\cdot\exp(-2\pi\, i\, \nu_0\, t)\cdot\exp(-t/2\, T_1)\cdot\Theta(t)$ is assumed.
The carrier frequency $\nu_0$ is given by the exciton energy $E_X=h\nu_0$ in the QD.
The wave packet is emitted at time $t=0$ as defined by the Heaviside function $\Theta(t)$.
Fourier transform (FT) $\mathbf{E}_{in}(t)\xrightarrow{\textrm{FT}}\mathbf{E}_{in}(\nu)$ of the temporal wave packet results in a Lorentzian spectral density $\mathbf{E}_{out}(\nu)$ with a Fourier-limited, natural homogeneous linewidth $\Delta\nu_{hom}= 1/(2\pi T_1)$.

The temporal wave packet at the end of the medium is calculated by inverse Fourier transform (iFT) of the output spectrum: $\mathbf{E}_{out}(\nu)\xrightarrow{\textrm{iFT}}\mathbf{E}_{in}(t)$.
The influence of spectral diffusion \cite{vural_two-photon_2018} on the QD emission frequency $\nu_0$ is taken into account, here, for the averaged intensity envelope
$\bar I_{out}(t)=\int_{-\infty}^\infty \textrm{iFT}\lbrace \mathbf{E}_{out}^\ast(\nu,\nu_0)\cdot\mathbf{E}_{out}(\nu,\nu_0)\cdot G(\nu_0,\Delta\nu_{inhom}) \, \rbrace \, \mathrm d\nu_0$
by integration over all possible output spectra $I_{out}(\nu,\nu_0)=\mathbf{E}_{out}^\ast(\nu,\nu_0)\cdot\mathbf{E}_{out}(\nu,\nu_0)$
with the inhomogeneous Gaussian broadening profile $G(\nu_0,\Delta\nu_{inhom})$ as weighting function of $\Delta\nu_{inhom}$ FWHM spectral width.
In case of spectral filtering with a Fabry-P\'erot resonator, a Lorentzian filter function $L(\nu,\nu_f,\Delta\nu_{FP})$ with a width of $\Delta\nu_{FP}$ is factored in: $\bar I(t)=\int_{-\infty}^\infty \textrm{iFT}\lbrace \mathbf{E}_{out}^\ast(\nu,\nu_0)\cdot\mathbf{E}_{out}(\nu,\nu_0)\cdot G(\nu_0,\Delta\nu_{inhom})\cdot L(\nu_0,\nu_f,\Delta\nu_{FP}) \rbrace \, \mathrm d\nu_0$.
As a last step, the timing jitter $\tau_{IRF}$ of the detection system is taken into account as the convolution of the calculated temporal profile of the broadened pulse with the Gaussian instrument response function: $\bar I_{IRF}(t)=\bar I_{out}(\tau)\ast G(t-\tau,\tau_{IRF})$.

\bibliography{qd_delay_refs.bib}

\section*{Acknowledgements}
This work was supported by the German Research Foundation (DFG) Collaborative Research Center (CRC) SFB 787 project C2, the German Federal Ministry of Education and Research (BMBF) project Q.Link-X, as well as the European Research Council (ERC) under the European Unions Horizon 2020 Research and Innovation Programme (SPQRel – Entanglement distribution via Semiconductor-Piezoelectric Quantum-Dot Relays, Grant Agreement No. 679183).
T.K. acknowledges funding by Rosa Luxemburg Foundation.

\section*{Author contributions statement}
T.K., J.W., A.A., A.W.S., and O.B. conceived the experiments. J.S.W, E.Z., R.T., A.R., and O.G.S. provided the quantum dot sample. T.K., J.W., and A.T. conducted the experiments. T.K. analysed the results and wrote the manuscript with subsequent contributions from J.W. and O.B. All authors contributed to the manuscript, reviewed and have approved submission of the final version of the manuscript.

\section*{Additional information}
\textbf{Competing interests:}
The authors declare no competing interests.

\end{document}